\documentclass{appolb}

\usepackage{graphicx}
\usepackage{wasysym}
\usepackage{amssymb}

\newcommand{\rin}{r_\mathrm{in}}
\newcommand{\rout}{r_\mathrm{out}}
\newcommand{\mc}{M_\mathrm{c}}
\newcommand{\md}{M_\mathrm{d}}

\begin{document}

\title{Geometry of Keplerian disk systems and bounds on masses of their components}
\author{Patryk Mach, Edward Malec and Micha\l~Pir\'og
\address{M. Smoluchowski Institute of Physics, Jagiellonian University, Reymonta 4, 30-059 Krak\'ow, Poland}}

\maketitle

\begin{abstract}
\tolerance=500
We investigate accreting disk systems with polytropic gas in Keplerian motion. Numerical data and partial analytic results show that the self-gravitation of the disk speeds up its rotation -- its rotational frequency is larger than that given by the well known strictly Keplerian formula that takes into account the central mass only. Thus determination of central mass in systems with massive disks requires great care -- the strictly Keplerian formula yields only an upper bound. The effect of self-gravity depends on geometric aspects of disk configurations. Disk systems with a small (circa $10^{-4}$) ratio of the innermost radius to the outermost disk radius have the central mass close to the upper limit, but if this ratio is of the order of unity then the central mass can be smaller by many orders of magnitude from this bound.
\end{abstract}

\section{Introduction}
\label{sect_introduction}

The literature on those accretion disk systems where the gas exhibits Keplerian motion $\omega^2 = \omega_0^2/r^3$ (here $\omega$ is the angular velocity of the gas, $r$ denotes the cylindrical radius, and $\omega_0$ is a constant) brings a confusing message. Usually the velocity of rotation of such disks is understood to be dominated by a heavy central mass $\mc$, that is the strictly Keplerian equality  $\mc = \omega_0^2/G$, where $G$ is the gravitational constant, holds true. The disk is assumed to be light. The analysis of the massive black hole in the masing nucleus of NGC 4258 exemplifies this interpretation \cite{Miyoshi1995} (Interestingly, modelling of this system confirms that the disk is light \cite{Mach}). On the other hand, there are sporadic observations in the literature that even systems with massive disks can conform to the Keplerian rotation curve \cite{Hashimoto1995}. In such a case the central mass cannot be deduced solely from the Keplerian frequency. Quite recently Hur\'{e} et al. \cite{Hure2011} investigated thin dust disks and, assuming a monomial density profile, arrived to a conclusion that the rotational Keplerian velocity gives only information on the enclosed mass of the system, i.e., a sum of the central mass $\mc$ and the mass of the disk $\md$.

In this paper we deal with self-gravitating barotropic disks in Keplerian motion. We show that, contrary to the folk belief, the Keplerian rotation curve in pure hydrodynamics does not exclude heavy disks. The angular frequency of the rotation curve does depend on the two masses, but they do not contribute additively. We start the discussion with analytic results  concerning configurations  with strictly (or almost strictly) Keplerian rotation law, that is where $\omega_0^2$ is equal (or  close)   to $G \mc$. This happens for elongated disks, for which the ratio of the innermost radius $\rin$ to the outermost one $\rout$ is small (or tends to zero), or in the test-fluid regime, that is when the gravity of the disk can be neglected.  

In numerical calculations we specialise to polytropic equations of state. The results show that the angular frequency $\omega_0$ satisfies $G \mc \le \omega_0^2 \le G(\mc + \md)$, but the precise value of $\omega_0$ depends on the geometry of the disk, in particular on the ratio $\rin/\rout$. If $\rin / \rout$ is small enough, then the central mass is reasonably well approximated by the Keplerian expression $\omega_0^2/G$. A significant deviation from this value can occur for elongated disks when their mass exceeds the central mass by a few orders of magnitude. In contrast to that, when  $\rin / \rout = 1/2$, the ratio $\omega_0^2/G$ yields a value close to the central mass $\mc$ only when in addition $\md \ll \mc$; in such a case the disk must be geometrically thin. In this specific example, obtained for the polytropic index $\Gamma = 5/3$, if $\md \approx \mc$, the disk becomes thick and $\omega_0^2/(G \mc) \approx 1.3$ instead of 1; the latter ratio grows rapidly with the further increase of $\md$.

Models of equilibrium toroidal or disk-like figures constitute a great simplification of the astrophysical reality. While such configurations appear in the late stages of current simulations of binary mergers, the theory of thick accretion disks would require more complex modelling. Nevertheless, we are convinced that a good understanding of the properties of self-gravitating equilibrium disk-like figures is a prerequisite for the construction of more realistic models (cf. \cite{frank_king_reine}). We return to this and connected issues in Sec.~\ref{summary}.

\section{Equations}

We consider a stationary, self-gravitating disk of gas rotating around a central point mass $\mc$. The gravitational potential of the system can be written as the superposition $\Phi = - G \mc/|\mathbf x| + \Phi_\mathrm{g}$, where the potential $\Phi_\mathrm{g}$ is due to the gravity of the disk. It satisfies the Poisson equation
\begin{equation}
\label{poisson}
\Delta \Phi_\mathrm{g} = 4 \pi G \rho,
\end{equation}
where $\rho$ denotes the mass density of the gas. Here $\mathbf x = (x, y, z)$ are the Cartesian coordinates originating at the central point-like mass. In the following we will also work in the cylindrical coordinates $(r,\phi,z)$.

Assume that the disk is axially symmetric and occupies a finite volume. The last demand imposes a condition onto equations of state, as discussed in Appendix A. For technical reasons we will also assume that the fluid region is connected, and that it is equatorially symmetric. The velocity of the gas is $\mathbf U = \omega(r,z) \partial_\phi$ or equivalently $\mathbf{U} = r \omega(r,z) \hat \mathbf{e}_\phi$, where $\hat \mathbf{e}_\phi$ denotes the unit azimuthal vector. A classical result due to Poincar\'{e} and Wavre states that for a barotropic equation of state $\omega$ depends only on the cylindrical radius $r$ \cite{Tassoul1978}. We will deal later with polytropic equations of state $p = K\rho^\Gamma$, where $p$ is the gas pressure, and $K$ and $\Gamma$ are constant. The Euler equations 
\[ \nabla p + \rho (\mathbf U \cdot \nabla) \mathbf U + \rho \nabla \Phi = 0 \]
can be integrated, yielding
\begin{equation}
\label{euler}
h + \Phi_\mathrm{c} + \Phi = C
\end{equation}
in the closure of a region  where $\rho $ is nonzero . Here $h$ denotes the specific enthalpy of the fluid: $dh = dp/\rho$, and
\[ \Phi_\mathrm{c} = - \int^r dr^\prime r^\prime \omega^2 \left( r^\prime \right) \]
is the centrifugal potential. The structure of the disk can be obtained from Eqs.~(\ref{poisson}) and (\ref{euler}) provided that the equation of state and the rotation law $\omega = \omega(r)$ are known. We would like to point out that the process of solving of Eq.~(\ref{euler})  consists in the simultaneous finding of the unknown functions and of the volume where the enthalpy $h$ is nonnegative.

\section{Keplerian rotation laws}
\label{sec_kepler}

In the rest of this paper we investigate systems with  the  Keplerian rotation laws 
\begin{equation}
\label{kepl}
\omega =  \frac{\omega_0}{r^{3/2}}.
\end{equation}
They are of special interest, because they agree with many observations. Almost all known stellar black holes with gaseous disks and a few of AGN's obey this law.  

Suppose now that in an accretion system the motion of a gaseous disk conforms to a Keplerian rotation law. The mass of gas (usually unknown to an observer) can impact the rotation curve. \textit{How legitimate is the (often made) guess that the motion is almost strictly Keplerian, $\omega_0^2 \approx GM_\mathrm{c}$, that allows one to infer the central mass from the observed rotation law?} Analytic answers can be given in a few special cases. \textit{i)} A thin gaseous Keplerian disk, with a mass negligibly small in comparison to the central mass $\mc$, exhibits strictly Keplerian motion, i.e., $\omega_0^2 = G \mc$. This is a well known fact, and we give in Sec.~3.1 a proof, only for the sake of completeness. Thus, in this case, one can obtain the value of the central mass simply by measuring the rotation frequency. \textit{ii)} For thick and heavy disks the situation is more complex. There is still a possibility to prove analytically the approximate equality $\omega_0^2 \approx GM_\mathrm{c}$, but only in special cases and under certain assumptions on disk's geometry. These results are discussed in Sec.~3.2 and 3.3.

In the generic case the presence of massive disks influences the rotation curve, or strictly saying, the frequency parameter. Our numerical analysis, reported in later sections, suggests that the difference $\omega_0^2 - GM_\mathrm{c}$ is strictly positive and increases with the increase of $\md / \mc$, where $\md \equiv \int d^3x \rho$ is the mass of the disk. A result of Hur\'{e} et al. \cite{Hure2011} (obtained for dust) suggests that the difference $\omega_0^2 - GM_\mathrm{c}$ should be proportional to the ratio $\md / \mc$. We show later that this is not true for polytropes.

The centrifugal potential for the rotation law (\ref{kepl}) can be written in the form $\Phi_\mathrm{c} = \omega_0^2/r$. Suppose that the disk is finite and extends from the innermost cylinder labelled by $\rin$ to the outermost cylinder at $r = r_\mathrm{out}$, and the innermost and outermost points of the disk are located on the equatorial plane. It follows from Eq.~(\ref{euler}) that
\begin{equation}
\label{omega_from_c}
\omega_0^2 = G \mc -  \rin \Phi_\mathrm{g}(\rin)  + \rin C  =  G \mc -  \rout \Phi_\mathrm{g}(\rout)  + \rout C,
\end{equation}
or
\begin{equation}
\label{omega}
\omega_0^2 = GM_\mathrm{c} + \frac{r_\mathrm{in} r_\mathrm{out} \left( \Phi_\mathrm{g}(r_\mathrm{out}) - \Phi_\mathrm{g}(r_\mathrm{in}) \right)}{r_\mathrm{out} - r_\mathrm{in}},
\end{equation}
because $h$ should vanish both for $\rin$ and $\rout$. Here $\Phi_\mathrm{g}(r_\mathrm{in}) = \Phi_\mathrm{g}(r=r_\mathrm{in},z=0)$ and $\Phi_\mathrm{g}(r_\mathrm{out}) = \Phi_\mathrm{g}(r=r_\mathrm{out},z=0)$. 

There are simple analytic arguments -- we discuss them in Sec.~3.4 -- showing that $\omega^2_0$ exceeds $G\mc$, when the constant $C \ge 0$. In the most of this paper, however, we deal with finite-volume configurations for which $C$ can be negative, and the analytic argument of Sec.~3.4 does not work (the relation between the equation of state, Keplerian rotation, and the sign of $C$ is discussed in Appendix A). Nevertheless, our numerical data reported in later sections do suggest that in fact $G\mc  \le \omega_0^2 \le G (\mc + \md)$. We state this as a conjecture: \textit{Barotropic Keplerian disks rotate with the angular velocity exceeding $(GM_\mathrm{c}/r^3)^{1/2}$, but smaller than $(G(\mc + \md)/r^3)^{1/2}$}.

\subsection{Test-fluid limit}

Massless Keplerian disks must be \textit{i)} strictly Keplerian and \textit{ii)} infinitely thin.

Indeed, choose disk configurations with fixed $\rin$ and $\rout$, but let the density $\rho_\mathrm{max} \to 0$. In this limit, the potential $\Phi_\mathrm{g}$ also vanishes ($\Phi_\mathrm{g}$ satisfies Eq.~(\ref{poisson}), and it is normalised to zero at infinity). Part \textit{i)} follows now from Eq.~(\ref{omega}), which yields $\omega_0^2 \to G\mc$.

To prove part \textit{ii)}, let us observe that Eq.~(\ref{omega_from_c}) yields $C = 0$ in the limit of $\Phi_\mathrm{g} \to 0$. In this case Eq.~(\ref{euler}) should be written as
\[ h + G \mc \left( \frac{1}{r} - \frac{1}{\sqrt{r^2 + z^2}} \right) = 0, \]
and this form admits (notice that $h \to 0$ when $\rho_\mathrm{max} \to 0$) only infinitely thin disks at $z = 0$.

\subsection{Disks with small inner radii}
\label{sec_limits}

Let us fix the maximum density within the disk, say $\rho = \rho_\mathrm{max}$, and let $\rout$ be also fixed. This choice is justified by the scaling symmetry of the equations (see Sec.~\ref{rescaling}). Then, it follows from Eq.~(\ref{omega}) that $\omega_0^2 \to G \mc$ as $\rin \to 0$. To show this, it suffices to note that $\Phi_\mathrm{g}$ is a bounded function on $\mathbb R^3$. To be more precise: $|\Phi_\mathrm{g}(\rout) - \Phi_\mathrm{g}(\rin)|$ must be finite when    $\rho_\mathrm{max}$ is fixed.

The above result can be intuitively understood, because for $r$ small enough, the gravitational potential is always dominated by the divergent term $-G \mc/|\mathbf x|$. In practice, if the ratio $\rin/\rout$ is sufficiently small, then the rotation curve of the gaseous disk should be influenced ``almost exclusively'' by the central mass. We show later, solving specific examples, that this is true.  

\subsection{Ring-like disks}

One can also show that $\omega_0^2 \to G \mc$ as $\rin \to  \rout$, that is when disk becomes ring-like with a small inner radius $(\rout - \rin)/2$. The proof proceeds according to the squeeze  theorem. For the sake of simplicity the second part of the reasoning (the bound from below) will be presented for polytropic equations of state, but the proof can be reformulated for general barotropes satisfying certain additional conditions.

The starting point is the virial relation formulated in Mach \cite{mach_virial} for the system consisting of a steady disk and a point mass. It yields
\begin{equation}
\label{virial_original}
\frac{1}{2} \int \rho \Phi_\mathrm{g} d^3 x - \int \rho \frac{G M_\mathrm{c}}{|\mathbf x|} d^3 x + \int \rho |\mathbf U|^2 d^3 x + 3 \int p d^3 x = 0.
\end{equation}
For Keplerian rotation this equation can be written as
\begin{equation}
\label{virial}
\frac{1}{2} \int \rho \Phi_\mathrm{g} d^3 x + \int \rho \left(\frac{\omega_0^2}{r} - \frac{G M_\mathrm{c}}{|\mathbf x|} \right) d^3 x + 3 \int p d^3 x = 0.
\end{equation}
Now, because $p \geq 0$ we have
\[  \int \rho \left(\frac{\omega_0^2}{r} - \frac{G M_\mathrm{c}}{|\mathbf x|} \right) d^3 x \leq - \frac{1}{2} \int \rho \Phi_\mathrm{g} d^3 x, \]
and, since $|\mathbf x| \geq r$,
\[ (\omega_0^2 - G M_\mathrm{c}) \int \frac{\rho}{r} d^3 x \leq - \frac{1}{2} \int \rho \Phi_\mathrm{g} d^3 x. \]
Let $\Phi_\mathrm{g,min}$ denote the minimum value of the potential $\Phi_\mathrm{g}$ in the region where $\rho \neq 0$. We have
\[ \frac{\omega_0^2 - G M_\mathrm{c}}{r_\mathrm{out}} \int \rho d^3 x \leq - \frac{1}{2} \Phi_\mathrm{g,min} \int \rho d^3 x,  \]
and thus
\[ \omega_0^2 - G M_\mathrm{c} \leq - \frac{1}{2} \Phi_\mathrm{g,min} r_\mathrm{out}. \]
For $r_\mathrm{in} \to r_\mathrm{out}$ the mass of the disk tends to zero (note that we keep $\rho_\mathrm{max}$ fixed, and the volume of the disk tends to zero), and so does the gravitational potential $\Phi_\mathrm{g,max}$. Thus, in the limit we have $\omega_0^2 - G M_\mathrm{c} \leq 0$.

On the other hand, the same formulation of the  virial theorem gives
\[ - 3 \int p d^3 x \leq \int \rho \left(\frac{\omega_0^2}{r} - \frac{G M_\mathrm{c}}{|\mathbf x|} \right) d^3 x \leq \left( \frac{\omega_0^2}{r_\mathrm{in}} - \frac{G M_\mathrm{c}}{r_\mathrm{out}} \right) \int \rho d^3 x, \]
which, for polytropic equations of state implies that
\[ - 3 K \rho_\mathrm{max}^{\Gamma - 1} \leq \frac{\omega_0^2}{r_\mathrm{in}} - \frac{G M_\mathrm{c}}{r_\mathrm{out}}. \]
From the integrated Euler equation we have
\[ \frac{K \Gamma}{\Gamma - 1} \rho_\mathrm{max}^{\Gamma - 1} + \Phi_\mathrm{g}(\hat r) + \frac{\omega_0^2}{\hat r} - \frac{G M_\mathrm{c}}{\hat r} = C = \Phi_\mathrm{g}(r_\mathrm{out}) + \frac{\omega_0^2}{r_\mathrm{out}} - \frac{G M_\mathrm{c}}{r_\mathrm{out}}, \]
where $\Phi_\mathrm{g}(\hat r) = \Phi_\mathrm{g}(r = \hat r, z = 0)$, and $\hat r$ denotes a cylindrical radius of the point where the density reaches its maximum. For simplicity we have assumed that the maximum lies on the equatorial plane. The second equality expresses the Euler equation at the outer boundary. Now, for $r_\mathrm{in} \to r_\mathrm{out}$ we have $\hat r \to r_\mathrm{out}$ as well. In this limit one obtains $K = 0$, so that $0 \leq \omega_0^2 - G M_\mathrm{c} \leq 0$. This concludes the proof.

\subsection{Infinite disks}

If disks are infinitely extended and posses a finite mass, then $C=0$. (An infinite extension imposes severe restrictions onto the allowed equations of state -- see also the discussion in Appendix A.) We claim that in this case $\omega_0^2 \ge GM_\mathrm{c}$.

Indeed, Eq.~(\ref{poisson}) yields, from the maximum principle \cite{Gilbarg1983}, that the potential $\Phi_\mathrm{g}$ is nonpositive everywhere and vanishes at spatial infinity. Thus Eq.~(\ref{omega_from_c}) implies that the rotational angular frequency must exceed the Keplerian value and $\omega_0^2 \ge GM_\mathrm{c}$.

The same argument would work for the constant $C>0$, but it is not clear whether such configurations exist.

\section{Rescaling of equations }
\label{rescaling}

It is convenient to transform Eqs.~(\ref{poisson}) and (\ref{euler}) into dimensionless forms.  The quantity $u = G r_\mathrm{out}^2 \rho_\mathrm{max}$ has the dimension of potentials. We define rescaled dimensionless potentials   $\tilde \Psi_\mathrm{c} = \Psi_\mathrm{c}/u$, $\tilde \Phi = \Phi/u$, $\tilde \Phi_\mathrm{g} = \Phi_\mathrm{g}/u$. The scaled enthalpy is given by $\tilde h = h/u$ while the transformed density reads   $\tilde \rho = \rho/\rho_\mathrm{max}$. New spatial coordinates  are defined as $\tilde {\mathbf x} = \mathbf x/r_\mathrm{out}$. Introducing $\tilde M_\mathrm{c} =G M_\mathrm{c}/(\rho_\mathrm{max} r_\mathrm{out}^3)$, we can split the new potentials as  $\tilde \Phi = - \tilde M_\mathrm{c}/|\tilde {\mathbf x}| + \tilde \Phi_\mathrm{g}$.   A similar    trick has been done in \cite{Blinnikov1975} and \cite{Hachisu1986}. The disk-related gravitational potential $\tilde \Phi_\mathrm{g}$ satisfies
\begin{equation}
\label{poisson_resc}
\tilde \Delta \tilde \Phi_\mathrm{g} = 4 \pi \tilde \rho,
\end{equation}
where $\tilde \Delta$ is the laplacian with respect to the new coordinates $\tilde {\mathbf x}$. The Euler equation (\ref{euler}) reads
\begin{equation}
\label{euler_resc}
\tilde h + \tilde \Phi_\mathrm{c} + \tilde \Phi = C.
\end{equation}
For the polytropic equations of state, the enthalpy $h$ can be expressed as $h = (K\Gamma/(\Gamma - 1)) \rho^{\Gamma - 1}$, and $\tilde h = (\tilde K\Gamma/(\Gamma - 1)) {\tilde \rho}^{\Gamma - 1}$, where constants $K$ and $\tilde K$ are related by $\tilde K = K \rho_\mathrm{max}^{\Gamma - 1}/u$. Finally, for the Keplerian rotation law we can define $\tilde \Phi_\mathrm{c} = \tilde \omega_0^2/\tilde r$. In this rescaled notation the Keplerian case  $\omega_0^2 = G M_\mathrm{c}$ corresponds to $\tilde \omega_0^2 = \tilde M_\mathrm{c}$.

In the new  variables, the set of parameters specifying the solution is reduced to $\Gamma$, $\tilde M_\mathrm{c}$ and $\tilde r_\mathrm{in} = r_\mathrm{in}/r_\mathrm{out}$. Taking large values of $\tilde M_\mathrm{c}$ corresponds to test-fluid solutions. Small values of $\tilde M_\mathrm{c}$ yield solutions with the massive disk.

\section{Numerical examples}

The  numerical method of this paper follows the classic Self-Consistent Field (SCF) scheme \cite{Ostriker1968, Clement1974, Blinnikov1975}. Eqs.~(\ref{poisson_resc}) and (\ref{euler_resc}) are solved iteratively on a two dimensional grid. In every iteration step the gravitational potential of the disk is found from Eq.~(\ref{poisson_resc}), assuming that the density is known. A new density distribution is then obtained directly from Eq.~(\ref{euler_resc}). Eq.~(\ref{poisson_resc}) is solved by the Green function formula, where the integral kernel is expanded in Legendre polynomials. In this approach the boundary conditions for the potential are automatically satisfied, and one can restrict the numerical grid to a region containing the disk. We would like to point out that this method requires renormalisation of  the constant $C$ appearing in (\ref{euler_resc}) and the polytropic constant $\tilde K$. They are adjusted in each iteration step in order to agree with the assumed parameters of the solution -- $\rin$, $\rout$ and the maximal value of $\tilde \rho$ equal to 1. The constant $\tilde \omega_0$ is computed from (\ref{omega}) at each  iteration step.

There exist modern numerical methods that are characterised by better convergence properties \cite{Eriguchi1985, Axenov1994}. Their implementation would probably be necessary  for non-Keplerian rotation laws. For Keplerian rotation laws numerical solutions converge   well even with the SCF method -- one can  obtain a solution for nearly any set of parameters. The drawback of the method is the need to employ a large number of Legendre polynomials for thin disks. One  is able,   using a parallel version of the code and optimisations described in \cite{Steinmetz},  to find accurate solutions using up to 400 of Legendre polynomials and   large grids (up to $5000 \times 5000$). Such resolutions are permitted because of the simplicity of the SCF method.

Our numerical solutions successfully passed a  virial test in the new formulation that includes a point mass \cite{mach_virial}. The virial theorem (\ref{virial_original}) can be written as
\[ E_\mathrm{pot} + 2 E_\mathrm{kin} + 2 E_\mathrm{therm} = 0, \]
where $E_\mathrm{pot} = \int d^3 x \rho \Phi_\mathrm{g}/2 - G \mc \int d^3 x \rho/|\mathbf x|$, $E_\mathrm{kin} = \int d^3 x \rho |\mathbf U|^2/2$, $E_\mathrm{therm} = \int d^3 x 3 p /2$. For the tested quantity we choose
\[ \epsilon_\mathrm{v} = |E_\mathrm{pot} + 2 E_\mathrm{kin} + 2 E_\mathrm{therm}|/|E_\mathrm{pot}|. \]
We have obtained a satisfactory accuracy, documented by the smallness of the ratio $\epsilon_\mathrm{v}$. The value of $\epsilon_\mathrm{v}$ depends on the resolution of the numerical grid. Appendix B shows that $\epsilon_\mathrm{v}$ decreases from the value of $10^{-4}$ to $10^{-8}$ with the finessing of the resolution.
\begin{figure}
\begin{center}
\includegraphics[width=10cm]{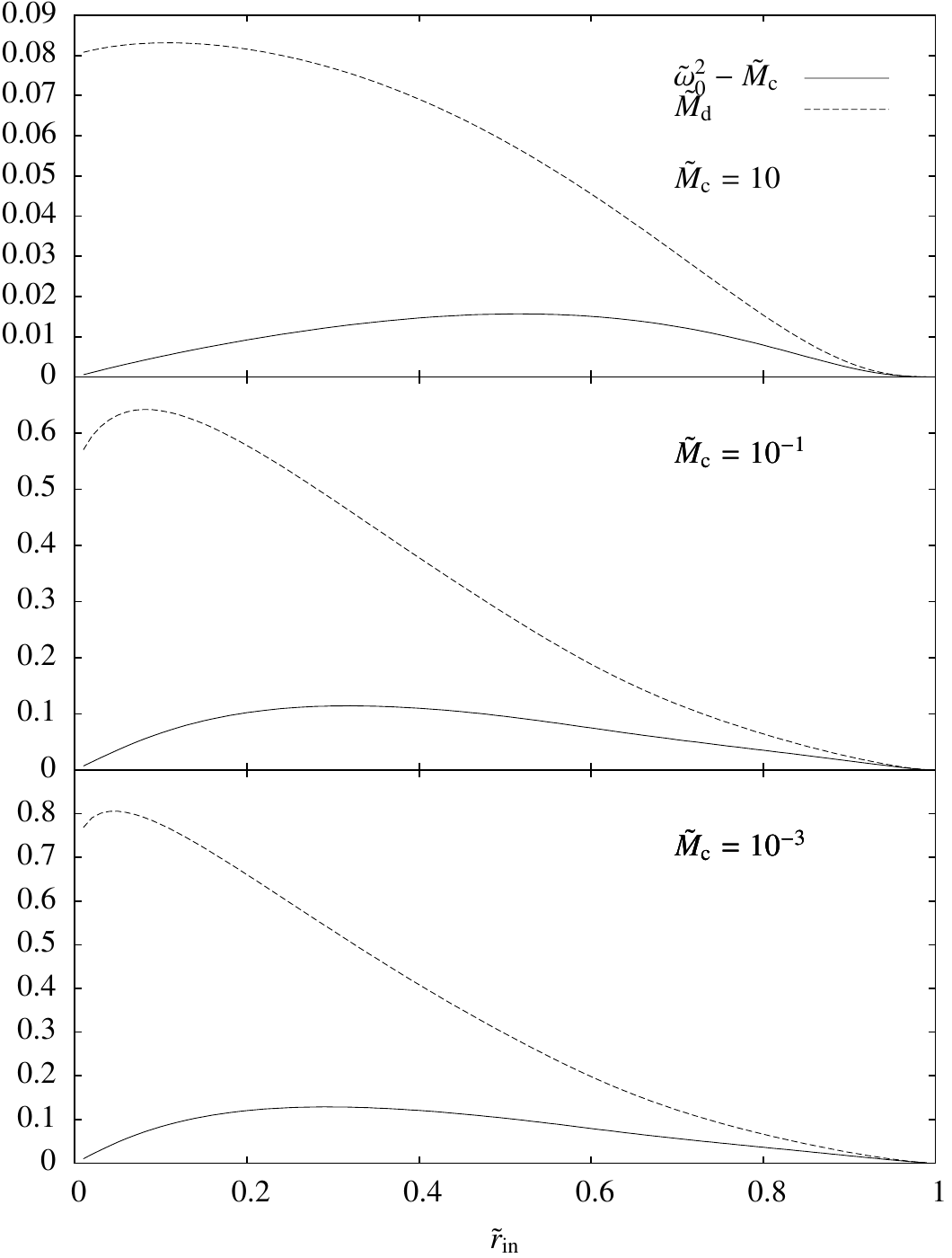}
\end{center}
\caption{\label{fig1} The plot shows $\tilde \omega_0^2 - \tilde \mc$ (solid line) and $\tilde \md$ (dotted line), as functions of $\rin$.  The graphs are obtained for $\Gamma = 5/3$, and different values of $\tilde \mc = 10, 10^{-1}$ and $10^{-3}$ respectively.}
\end{figure}

\begin{figure}
\begin{center}
\includegraphics[width=10cm]{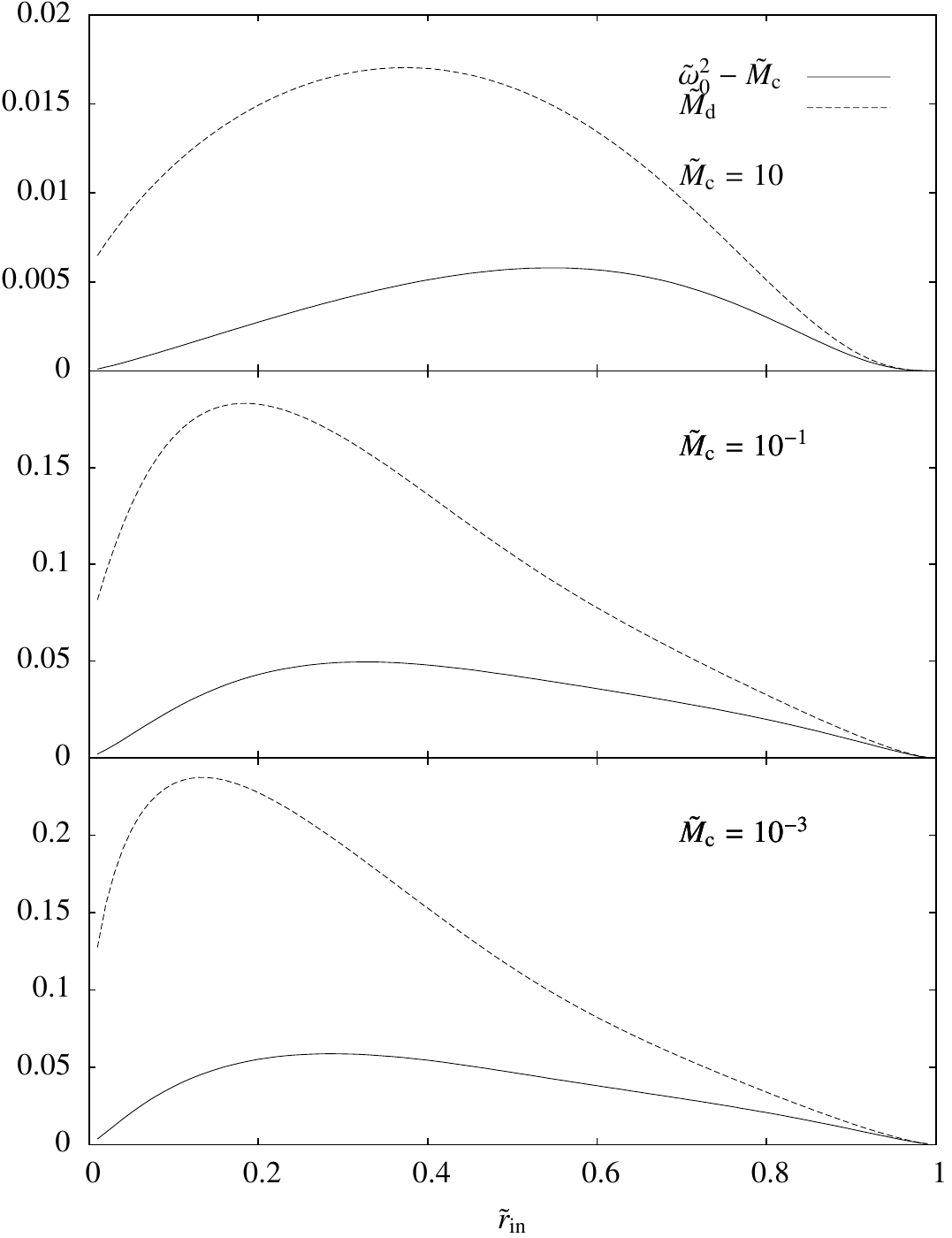}
\end{center}
\caption{\label{fig2} The same as in Fig.~\ref{fig1}, but for $\Gamma = 4/3$.}
\end{figure}

\begin{figure}
\begin{center}
\includegraphics[width=10cm]{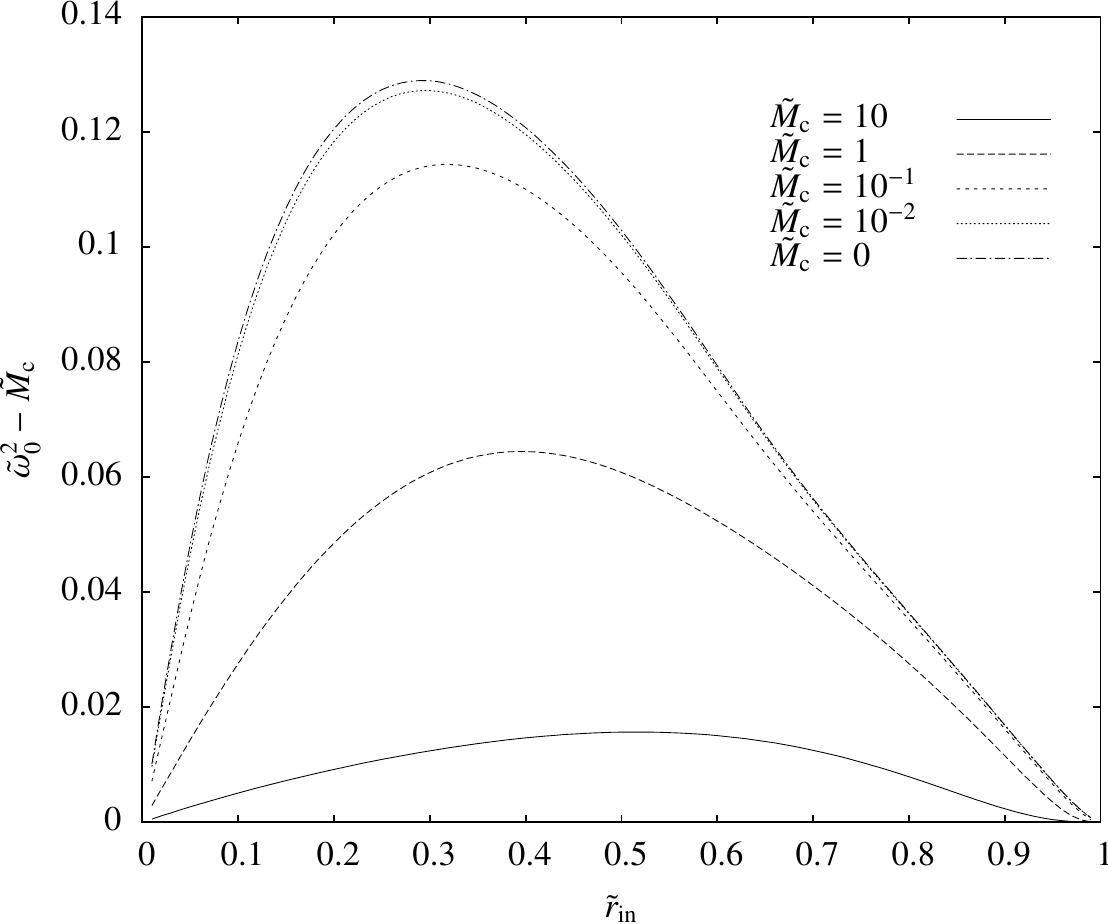}
\end{center}
\caption{\label{fig3} Dependence of $\tilde \omega_0^2 - \tilde \mc$ on $\tilde \rin$ for different values of $\tilde \mc$ and $\Gamma = 5/3$.}
\end{figure}

\begin{figure}
\begin{center}
\includegraphics[width=10cm]{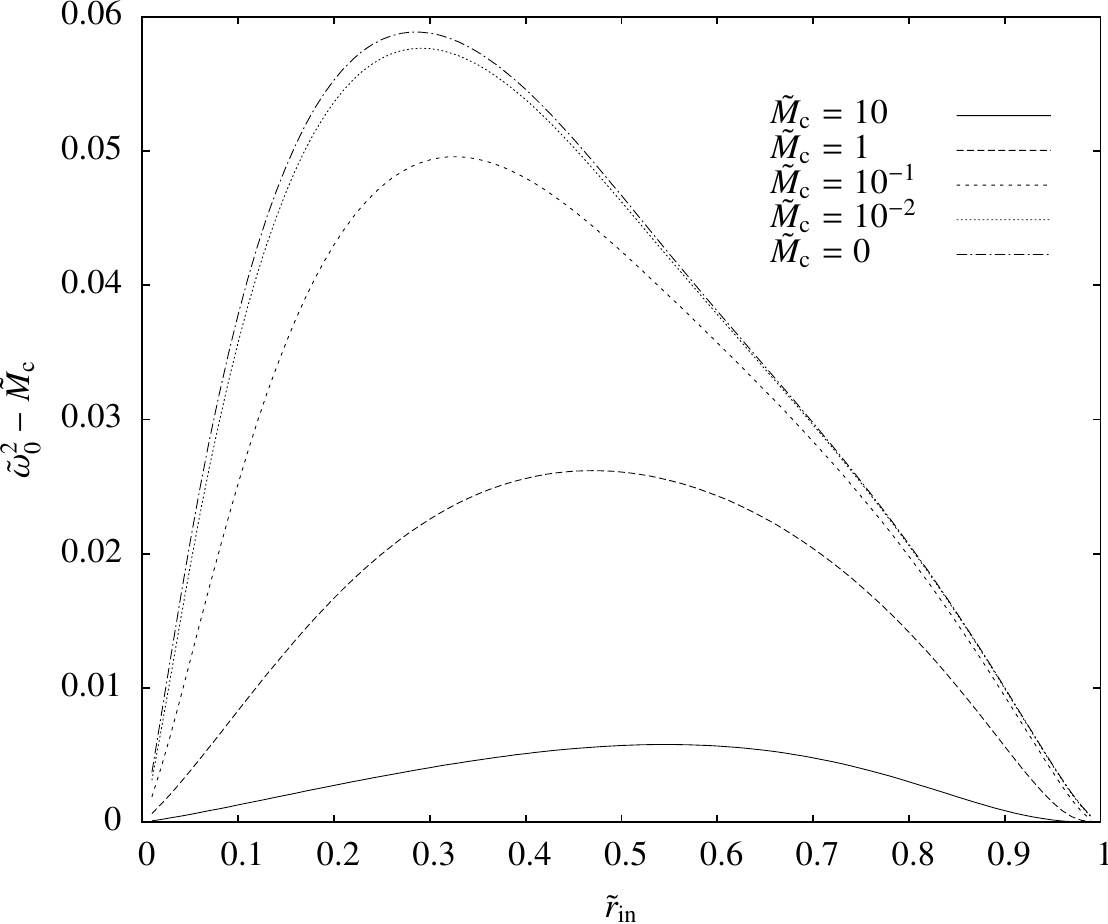}
\end{center}
\caption{\label{fig4} The same as in Fig.~\ref{fig3}, but for $\Gamma = 4/3$.}
\end{figure}

\begin{figure}
\begin{center}
\includegraphics[width=10cm]{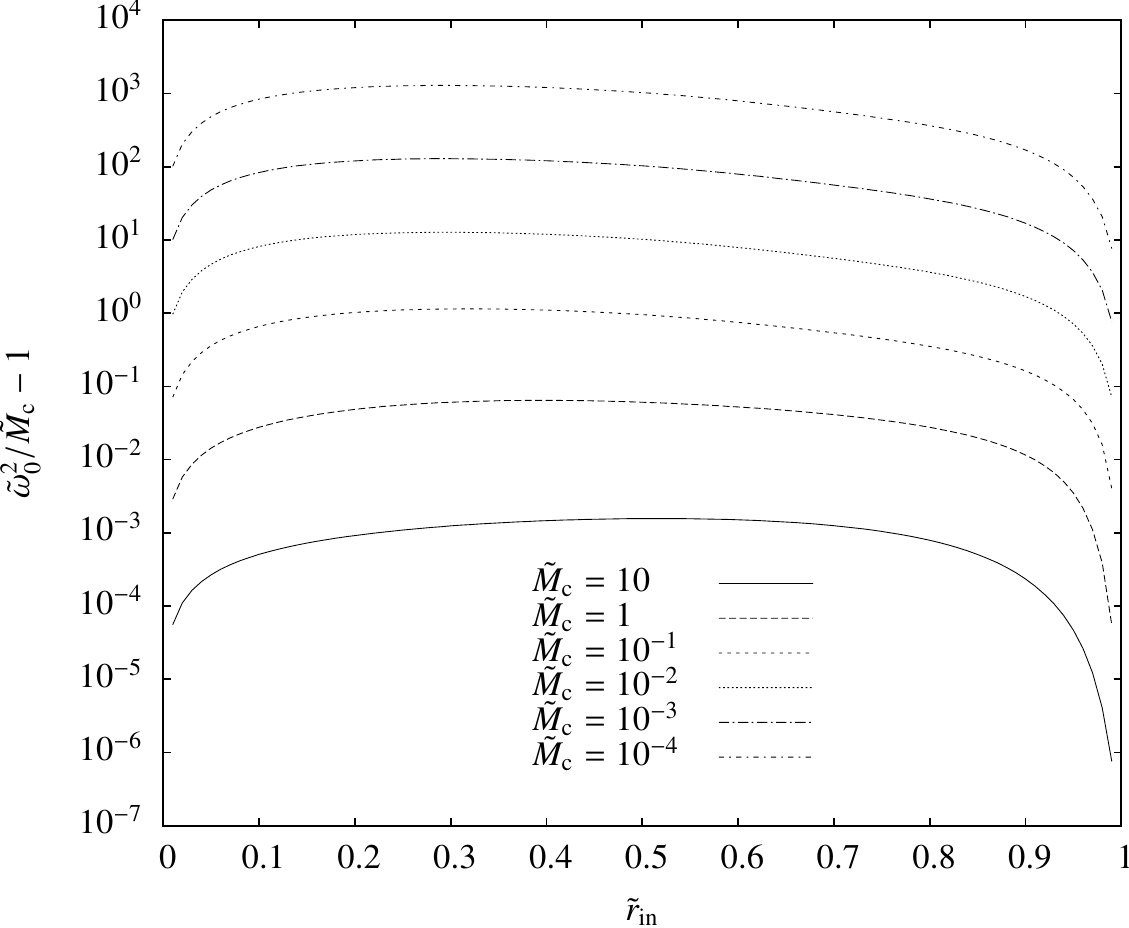}
\end{center}
\caption{\label{fig5} Dependence of $\tilde \omega_0^2/\tilde \mc - 1$ on $\tilde \rin$ for different values of $\tilde \mc$ and $\Gamma = 5/3$.}
\end{figure}

\begin{figure}
\begin{center}
\includegraphics[width=10cm]{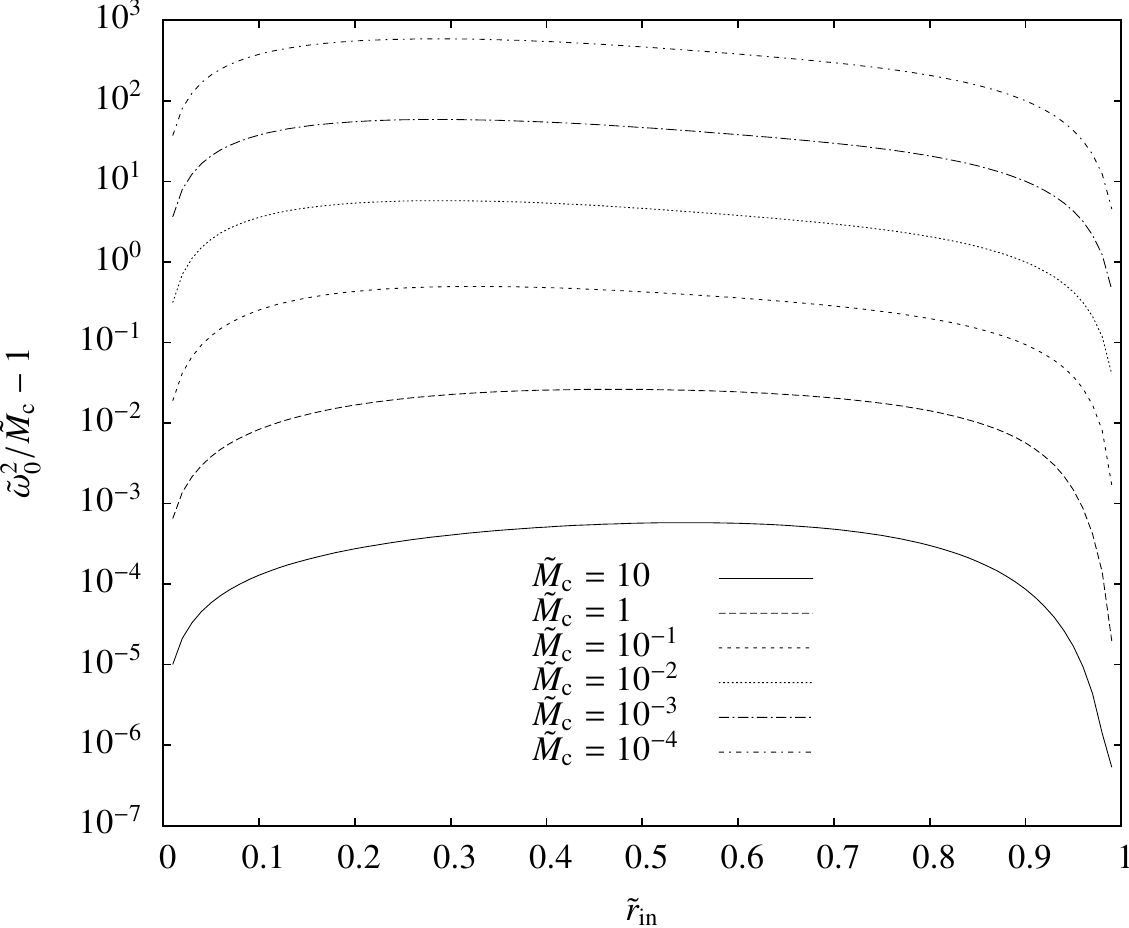}
\end{center}
\caption{\label{fig6} The same as in Fig.~\ref{fig5}, but for $\Gamma = 4/3$.}
\end{figure}

\begin{figure}
\begin{center}
\includegraphics[width=10cm]{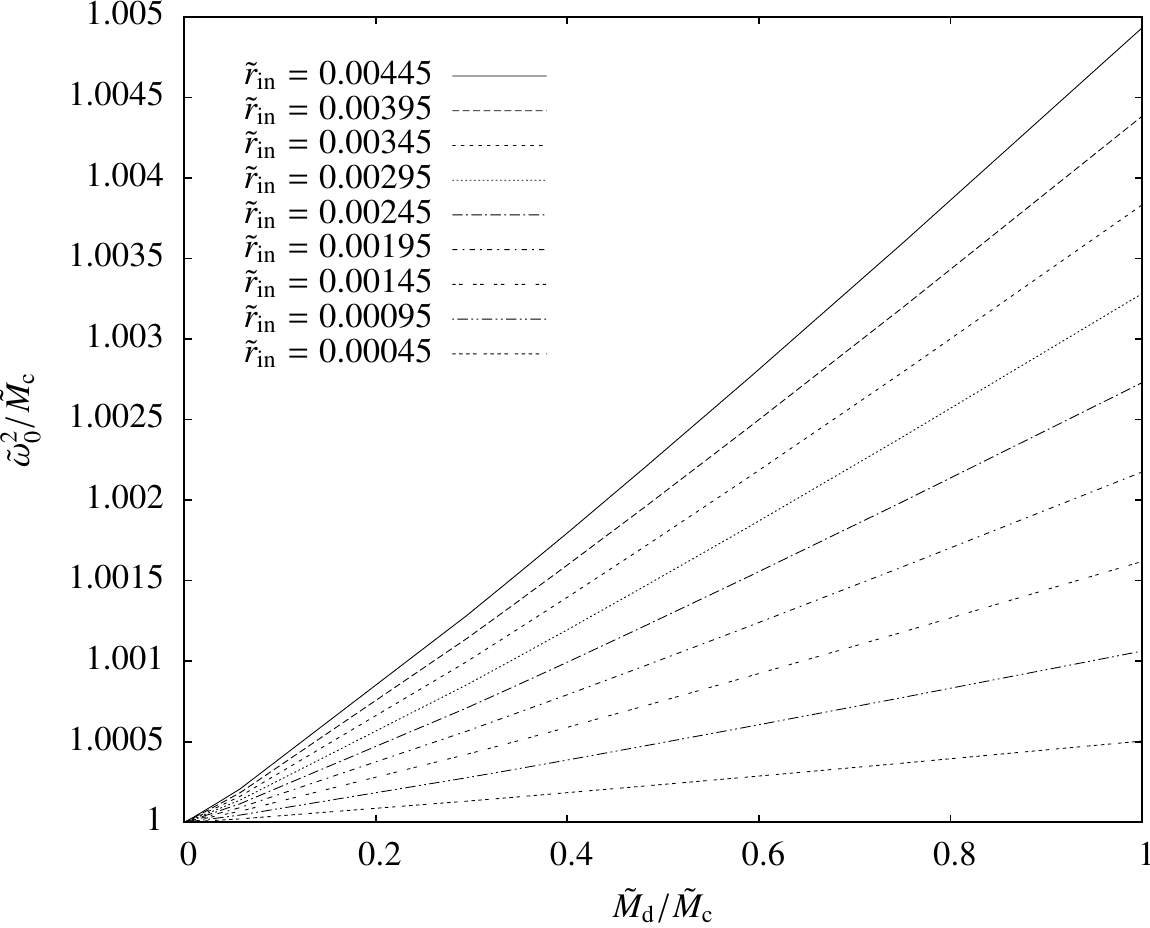}
\end{center}
\caption{\label{fig7} Dependence of $\tilde \omega_0^2/\tilde \mc$ on $\tilde \md/\tilde \mc$ for different values of $\tilde \rin$ and $\Gamma = 5/3$.}
\end{figure}

\begin{figure}
\begin{center}
\includegraphics[width=10cm]{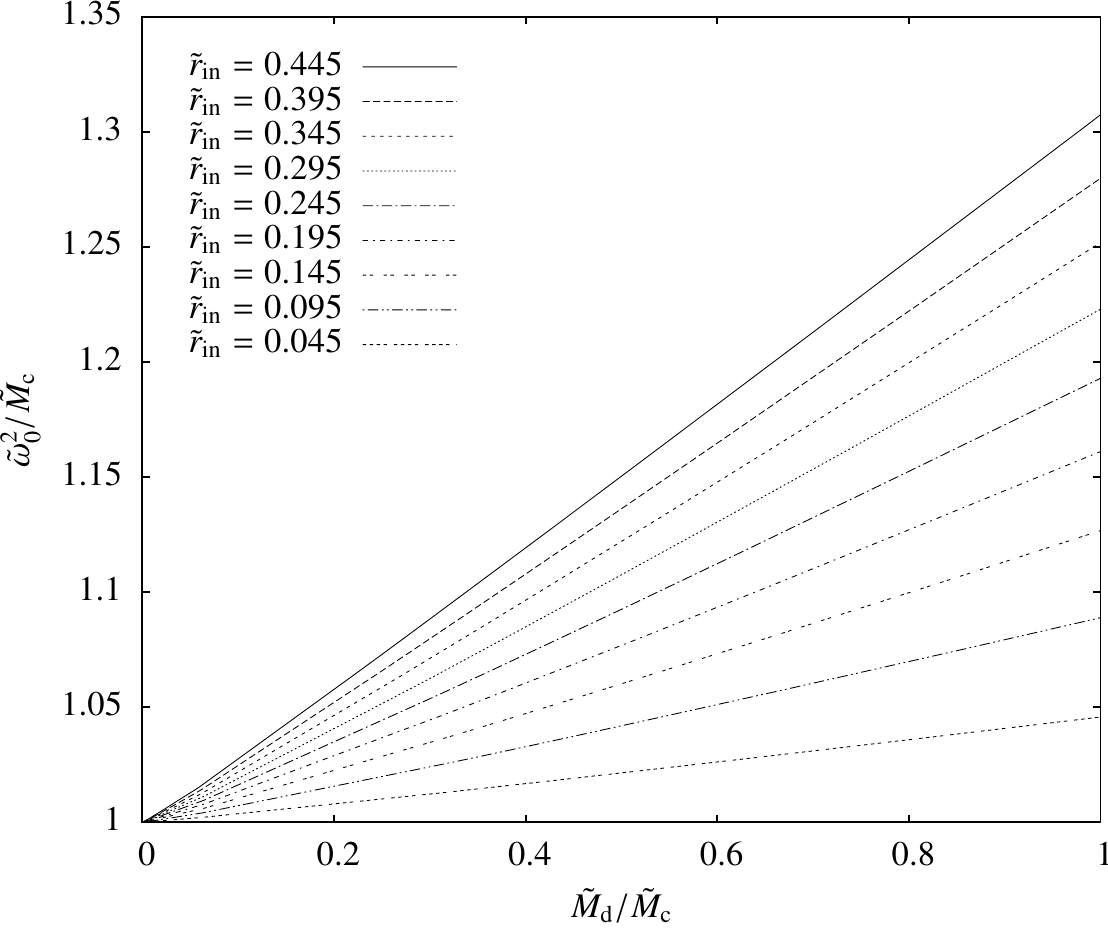}
\end{center}
\caption{\label{fig8} Dependence of $\tilde \omega_0^2/\tilde \mc$ on $\tilde \md/\tilde \mc$ for different values of $\tilde \rin$ and $\Gamma = 5/3$.}
\end{figure}

\begin{figure}
\begin{center}
\includegraphics[width=10cm]{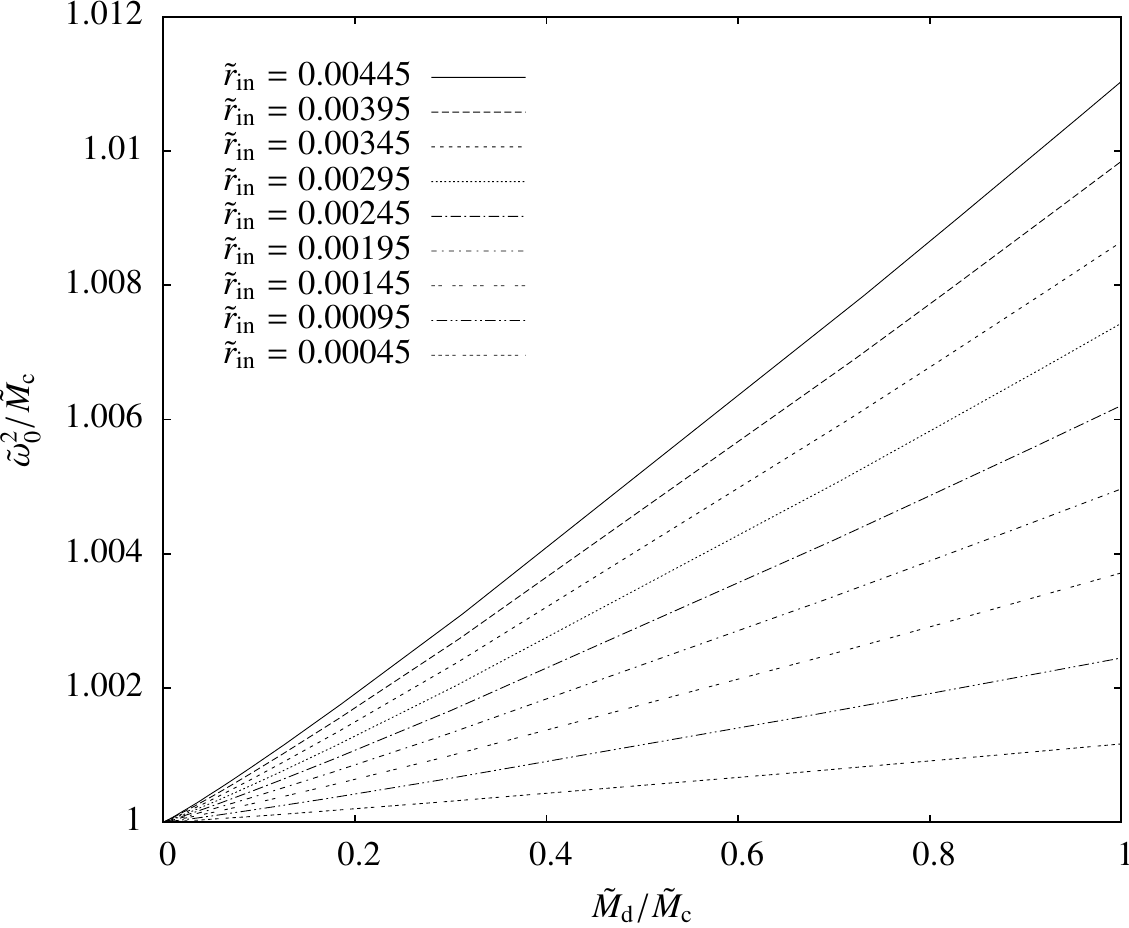}
\end{center}
\caption{\label{fig9} The same as in Fig.~\ref{fig7}, but for $\Gamma = 4/3$.}
\end{figure}

\begin{figure}
\begin{center}
\includegraphics[width=10cm]{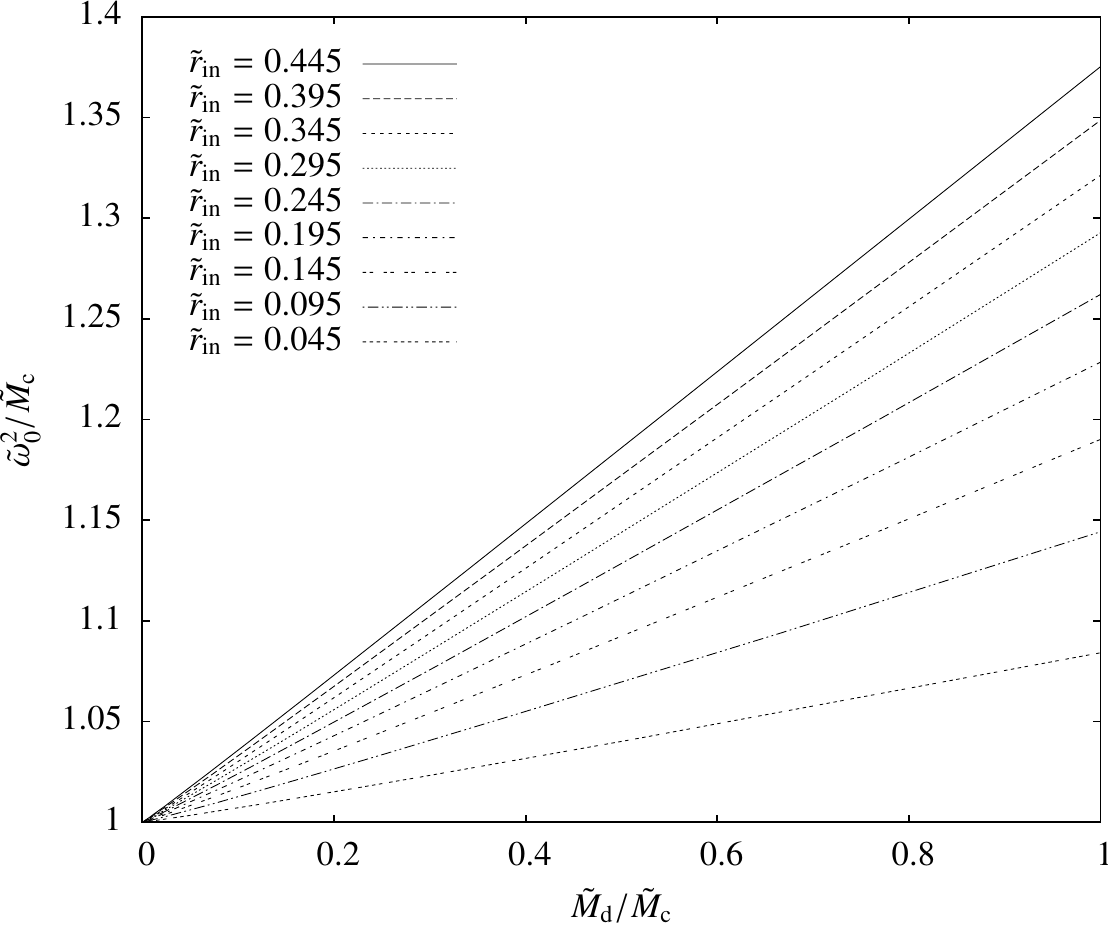}
\end{center}
\caption{\label{fig10} The same as in Fig.~\ref{fig8}, but for $\Gamma = 4/3$.}
\end{figure}

We performed a few thousands runs for polytropic disks, with two different polytropic indices  $\Gamma = 5/3$ and  $\Gamma = 4/3$. The results that we would like to discuss first are presented on Figs.~\ref{fig1}--\ref{fig6}. Figures \ref{fig1} and \ref{fig2} show the graphs of $\omega_0^2 - \tilde \mc$ and $\tilde \md$ versus $\tilde \rin$ for a couple of different values of $\tilde \mc$. They confirm the conjecture stated in Sec.~\ref{sec_kepler}, which in the rescaled variables can be written as $\tilde \mc \le \tilde \omega_0^2 \le \tilde \mc + \tilde \md$  or  $0 \le \tilde \omega_0^2 - \tilde \mc \le \tilde \md$. It is also clear that $\omega_0^2 \to \tilde \mc$ for $\tilde \rin \to 0$ and $\tilde \rin \to 1$, as it was proven in Sec.~\ref{sec_limits}.

Graphs of $\omega_0^2 - \tilde \mc$ for different values of $\tilde \mc$ and different polytropic exponents $\Gamma$ are shown on Figs.~\ref{fig3} and \ref{fig4}. We observe that this difference $\omega_0^2 - \tilde \mc$, treated as a function of $\tilde \rin$, tends to zero for $\tilde \mc \to \infty$ (test-fluid limit), and to a fixed function $\tilde \omega_0^2(\tilde \rin, \tilde \mc = 0)$ for $\tilde \mc \to 0$. The solution with $\tilde \mc = 0$ can be obtained numerically, and $\tilde \omega_0^2(\tilde \rin, \tilde \mc = 0)$ has a maximum of the order of 0.13 for $\Gamma = 5/3$ and 0.06 for  $\Gamma = 4/3$. That, in turn, yields upper estimates for the ratio $\tilde \omega_0^2/\tilde \mc$, namely
\begin{equation}
\label{upper_bound_53}
\frac{\tilde \omega_0^2}{\tilde \mc} \le 1 + \frac{0.13}{\tilde \mc} \;\;\; \mathrm{for} \;\;\; \Gamma = \frac{5}{3},
\end{equation}
and
\begin{equation}
\label{upper_bound_43}
\frac{\tilde \omega_0^2}{\tilde \mc} \le 1 + \frac{0.06}{\tilde \mc} \;\;\; \mathrm{for} \;\;\; \Gamma = \frac{4}{3}.
\end{equation}
 For the reasons of clarity, the values of $\omega_0^2/\tilde \mc - 1$ are also plotted in Figs.~\ref{fig5} and \ref{fig6}. The maximum value of the rescaled mass of the disk $\tilde \md$ is attained for $\tilde \mc = 0$. It is of the order of unity for $\Gamma = 5/3$ and even smaller for $\Gamma = 4/3$. Thus, it is clear from Eqs.~(\ref{upper_bound_53}) and (\ref{upper_bound_43}) that the central mass $\tilde \mc$ influences the angular velocity $\tilde \omega$ much stronger than $\tilde \md$, or, to put it differently, $\tilde \mc$ enters the expression for $\tilde \omega$ with a larger weight than $\tilde \md$.

Numerically obtained solutions correspond to thickest disks when $\tilde \mc = 0$, and become infinitely thin for $\tilde \mc \to \infty$, that is in the test-fluid limit. For a fixed value of $\tilde \rin$, the relative height of the disk $H/(\rin - \rout)$ (we assume that the disk extends from $z = -H$ to $z = H$) decreases monotonically with $\tilde \mc$. Thus, in principle, geometrical parameters such as the ratio of $\rin/\rout$ and the relative height of the disk should allow one to identify a particular solution and to obtain values of $\tilde \omega_0^2/\tilde \mc$ and $\tilde \md/\tilde \mc$, provided that the polytropic exponent is known.

Figures \ref{fig7}--\ref{fig10} show the dependence of $\tilde \omega_0^2/\tilde \mc$ on the mass ratio $\tilde \md / \tilde \mc$. We focus on the physically interesting range of parameters $0 < \tilde \md / \tilde \mc < 1$ and $0 < \tilde \rin < 1/2$. Results obtained for $\Gamma = 5/3$ are shown in Figs.~\ref{fig7} and \ref{fig8}; the data for $\Gamma = 4/3$ are depicted in Figs.~\ref{fig9} and \ref{fig10}. These graphs confirm quantitatively what was anticipated in Sec.~\ref{sect_introduction}. For small values of $\tilde \rin$ (elongated disks) the motion is strictly Keplerian with $\tilde \omega_0^2 \approx \tilde \mc$, even if the mass of the disk is comparable to the central mass. Faster rotation can occur for $\tilde \md \approx \tilde \mc$ only if in addition $\tilde \rin$ is large, but even for $\tilde \rin = 1/2$ we get $\tilde \omega_0^2 \approx 1.3$ for $\Gamma = 5/3$ and $\tilde \omega_0^2 \approx 1.4$ for $\Gamma = 4/3$. In general these results weakly depend on the polytropic index $\Gamma$.

\section{Summary}
\label{summary}

We show explicitly, by  analysing thousands of solutions of polytropic hydrodynamic models, that the self-gravity of a gaseous heavy disk  orbiting a central mass can  be compatible with the Keplerian motion.  Nevertheless  the  masses of the two components  can be resolved, if in addition to the rotation velocity two geometric pieces of information are given, namely the ratio of the innermost radius $\rin$ to the outermost disk radius $\rout$ and its relative maximal height $H/(\rout - \rin)$. 

How legitimate is the (often made) guess that rotation is strictly Keplerian,  $\omega_0^2 \approx GM_\mathrm{c}$, which allows one to infer the central mass from the observed rotation law?  A result of Hur\'{e} et al. \cite{Hure2011} (obtained for dust)  suggests that  the error implied by this identification  should be proportional to the ratio of $\md / \mc$, where $\md$ is the mass of the disk.  We show that this is not true for polytropes.
Our numerical results suggest that polytropic Keplerian disks rotate with the angular velocity exceeding $(GM_\mathrm{c}/r^3)^{1/2}$, but smaller than $(G(\mc + \md)/r^3)^{1/2}$. The lower bound in these inequalities can be obtained   analytically for special configurations, reported in Sec. 3. 

Presented results are based on hydrostationary modelling. Situation can be different for systems strongly influenced by physical factors that were not discussed here. For instance, models of strongly radiating thick toroids considered in \cite{Hashimoto1995} are not only Keplerian, but strictly Keplerian with $\omega_0^2 = G\mc$, although those disks are massive, and the value of $\rin/\rout$ is not small. In other words, very strong radiation can invalidate conclusions concerning hydrostationary systems. On the other hand, rotation velocity of weakly radiating disks analysed in \cite{mach_malec_2012} is consistent with the results of this paper.

This analysis is entirely Newtonian, and therefore it should be repeated in the general-relativistic context. We do not expect essential changes in those systems, where the inner disk boundary is well separated from a black hole horizon, $\rin \gg 6 \mc$. This issue -- and further related questions -- deserves separate studies.

\section*{Acknowledgements}

The research was carried out with the supercomputer ``Deszno'' purchased thanks to the financial support of the European Regional Development Fund in the framework of the Polish Innovation Economy Operational Program (contract no. POIG. 02.01.00-12-023/08).

\section*{Appendix A}

We derive in  this appendix various criteria -- restrictions on equations of state -- that guarantee finite disk  configurations. The configuration is called finite if it can be contained in a compact subset of $\mathbb R^3$. We redo the proof, since the existing versions \cite{mach_simon} do not include the case with the singular central potential $- G\mc/|\mathbf{x}|$.

We start with an observation that the inifiniteness in the axial direction $z$ is incompatible with the cylindrical Keplerian rotation law (any cylindrical rotation in fact). This can be demonstrated as follows. Assume that the fluid extends to infinity along $z$ direction for two different constant radii $r_1$ and $r_2$. Then from Eq.~(\ref{euler}) we have
\[  \lim_{z \to \infty, \; r = r_1} (h + \Phi + \Phi_\mathrm{c}) = \Phi_\mathrm{c}(r_1) = C, \]
and the same applies to $r = r_2$. Thus, we would conclude that $\Phi_\mathrm{c} (r_1) = \Phi_\mathrm{c} (r_2) = C$, which contradicts the assumption that the motion is Keplerian.

In principle the disk could still extend to infinity at the equatorial plane. Assume now that the disk has a finite mass. Then the only value of constant $C$ appearing in Eq.~(\ref{euler}) allowing for an infinite configuration is $C = 0$ (note that the Keplerian centrifugal potential is normalised to zero at $r \to \infty$). This situation can be excluded for some cases with the help of the virial theorem (\ref{virial}).

The first necessary step is to establish suitable function spaces for the density, and the pressure, that include infinitely extended fluids, and for which the virial theorem can still be proven. This can be done in terms of weighted Sobolev spaces. Details (which we omit) can be found in \cite{mach_simon}. Apart from the suitable falloff behaviour we also assume that the support of the density is a connected set.

Multiplying the integrated Euler equation by $\rho$ and integrating over $\mathbb R^3$ we get
\[ C \int \rho d^3 x = \int \rho h d^3 x + \int \rho \Phi_\mathrm{g} d^3 x + \int \rho \left( \frac{\omega_0^2}{r} - \frac{G M_\mathrm{c}}{|\mathbf x|} \right). \]
Combining this equation with the virial theorem (\ref{virial}) one obtains
\[ C \int \rho d^3 x = \frac{1}{2} \int \rho \Phi_\mathrm{g} d^3 x + \int (\rho h - 3p) d^3 x. \]
Now, for those barotropic equations of state for which $\rho h - 3 p \le 0$ everywhere, we can conclude that $C$ is strictly negative, and hence the fluid must be finite. For polytropic equations of state we have $\rho h - 3p = ((3 - 2 \Gamma)/(\Gamma - 1))p$, and the finiteness is guaranteed for $\Gamma \ge 3/2$.

The condition on the equation of state can be relaxed on the expense of additional assumptions on the rotation. Making use of the virial theorem we can also write
\[ C \int \rho d^3 x = \int (\rho h - 6 p) d^3 x - \int \rho \left( \frac{\omega_0^2}{r} - \frac{G M_\mathrm{c}}{|\mathbf{x}|} \right) d^3 x. \]
If $\omega_0^2/r - G M_\mathrm{c}/|\mathbf x| \ge 0$ everywhere, it is enough to demand that $\rho h - 6 p < 0$ in order to conclude that the fluid must be finite. For polytropes the later condition simply states that $\Gamma > 6/5$.

\section*{Appendix B}

Below we report results of the convergence tests for typical disk configurations obtained with our SCF numerical scheme. In our implementation, numerical precision is controled by the grid resolution, the maximum number of Legendre polynomials used in the angular expansion $L$, and a value of the maximal difference between density distributions obtained in the last two consecutive iterations $\tilde \rho_\mathrm{tol}$. (In each iteration we compute the quantity $\tilde \rho_\mathrm{err} = \max_{i,j} |\tilde \rho^{(k+1)}_{i,j} - \tilde \rho^{(k)}_{i,j}|$. Here index $k$ numbers subsequent iterations; indices $i$ and $j$ refer to different grid nodes. The iteration procedure is stopped, when $\tilde \rho_\mathrm{err} \le \tilde \rho_\mathrm{tol}$.)

Table \ref{tab1} shows the dependence of results ($\tilde \omega_0$ and the virial parameter $\epsilon_\mathbf{v}$) on the grid resolution, the maximum number of Legendre polynomials $L$, and $\tilde \rho_\mathrm{tol}$. For this test we have assumed $\tilde \mc = 1$, $\tilde \rin = 0.3 $, $\Gamma = 5/3$. The corresponding mass of the disk yields $\tilde \md \approx 0.29$.

\begin{table}
\caption{\label{tab1} Typical dependence of the results on the grid resolution, the maximum number of the Legendre polynomials $L$, and the tolerance coefficient $\tilde \rho_\mathrm{tol}$.}
\begin{tabular}{@{}ccccc}
\hline
Resolution & $L$ & $\tilde \rho_\mathrm{tol}$ & $\tilde \omega_0$ & $\epsilon_\mathrm{v}$ \\
\hline
100  $\times$ 100  & 200 & $10^{-5}$ & 1.029945714 & $2.22 \times 10^{-5}$\\
200  $\times$ 200  & 200 & $10^{-5}$ & 1.029948775 & $5.49 \times 10^{-6}$\\
400  $\times$ 400  & 200 & $10^{-5}$ & 1.029949540 & $1.30 \times 10^{-6}$\\
800  $\times$ 800  & 200 & $10^{-5}$ & 1.029949076 & $2.45 \times 10^{-7}$\\
1600 $\times$ 1600 & 200 & $10^{-5}$ & 1.029949123 & $1.67 \times 10^{-8}$\\
3000 $\times$ 3000 & 200 & $10^{-5}$ & 1.029949140 & $7.92 \times 10^{-8}$\\
5000 $\times$ 5000 & 200 & $10^{-5}$ & 1.029949137 & $9.51 \times 10^{-8}$\\
\hline
100  $\times$ 100  & 200 & $10^{-6}$ & 1.029945714 & $2.22 \times 10^{-5}$\\
200  $\times$ 200  & 200 & $10^{-6}$ & 1.029948775 & $5.49 \times 10^{-6}$\\
400  $\times$ 400  & 200 & $10^{-6}$ & 1.029949540 & $1.30 \times 10^{-6}$\\
800  $\times$ 800  & 200 & $10^{-6}$ & 1.029949076 & $2.45 \times 10^{-7}$\\
1600 $\times$ 1600 & 200 & $10^{-6}$ & 1.029949123 & $1.67 \times 10^{-8}$\\
3000 $\times$ 3000 & 200 & $10^{-6}$ & 1.029948709 & $2.54 \times 10^{-8}$\\
5000 $\times$ 5000 & 200 & $10^{-6}$ & 1.029948705 & $4.13 \times 10^{-8}$\\
\hline
2400 $\times$ 2400 & 100 & $10^{-5}$ & 1.029948963 & $7.35 \times 10^{-7}$\\
2400 $\times$ 2400 & 200 & $10^{-5}$ & 1.029949132 & $6.53 \times 10^{-7}$\\
2400 $\times$ 2400 & 300 & $10^{-5}$ & 1.029949145 & $6.47 \times 10^{-7}$\\
2400 $\times$ 2400 & 400 & $10^{-5}$ & 1.029949147 & $6.46 \times 10^{-7}$\\
2400 $\times$ 2400 & 500 & $10^{-5}$ & 1.029949148 & $6.45 \times 10^{-7}$\\
\hline
\end{tabular}
\end{table}

In addition, Tables \ref{tab2} and \ref{tab5} illustrate the dependence of the convergence properties on the value of $\tilde \mc$ (or, equivalently, the ratio  of $\mc/\md$). Data listed in these tables were obtained for $L=200$, $\Gamma = 5/3$, $\tilde \rin = 0.3 $, and $\tilde \rho_\mathrm{tol} = 10^{-5}$. Table \ref{tab2} corresponds to $\tilde \mc = 0$ (in this case $\tilde \md \approx 0.49$). In Table \ref{tab5} we assumed $\tilde \mc = 10$ ($\tilde \md \approx 1.4 \times 10^{-2}$).

\begin{table}
\caption{\label{tab2} Dependence of the results on the resolution of the grid for a massive disks with $\tilde \mc = 0$. Here $\tilde \rho_\mathrm{tol} = 10^{-5}$ and $L=200$.}
\begin{tabular}{@{}ccc}
\hline
Resolution & $\tilde \omega_0$ & $\epsilon_\mathrm{v}$ \\
\hline
100  $\times$ 100  & 0.339018059 & $1.32 \times 10^{-4}$\\
200  $\times$ 200  & 0.339031477 & $3.67 \times 10^{-5}$\\
400  $\times$ 400  & 0.339034828 & $1.29 \times 10^{-5}$\\
800  $\times$ 800  & 0.339035663 & $6.99 \times 10^{-6}$\\
1600 $\times$ 1600 & 0.339035431 & $5.51 \times 10^{-6}$\\
3000 $\times$ 3000 & 0.339035495 & $5.15 \times 10^{-6}$\\
5000 $\times$ 5000 & 0.339035486 & $5.06 \times 10^{-6}$\\

\hline
\end{tabular}
\end{table}

\begin{table}
\caption{\label{tab5} Dependence of the results on the resolution of the grid for a light disk with $\tilde \mc = 10 $. Here $\tilde \rho_\mathrm{tol} = 10^{-5}$ and $L=200$.}
\begin{tabular}{@{}ccc}
\hline
Resolution & $\tilde \omega_0$ & $\epsilon_\mathrm{v}$ \\
\hline
100  $\times$ 100  & 3.164121422 & $1.14 \times 10^{-6}$\\
200  $\times$ 200  & 3.164121530 & $2.32 \times 10^{-7}$\\
400  $\times$ 400  & 3.164121439 & $1.25 \times 10^{-9}$\\
800  $\times$ 800  & 3.164121436 & $5.66 \times 10^{-8}$\\
1600 $\times$ 1600 & 3.164121424 & $7.12 \times 10^{-8}$\\
3000 $\times$ 3000 & 3.164121425 & $7.46 \times 10^{-8}$\\
5000 $\times$ 5000 & 3.164121425 & $7.55 \times 10^{-8}$\\
\hline
\end{tabular}
\end{table}

Notice that the obtained value of the virial test parameter depends mainly on the resolution of the numerical grid. Values obtained in this paper agree with those reported by Axenov \& Blinnikov \cite{Axenov1994}, who were also testing their implementation on high-resolution grids.

\end{document}